\documentclass[12pt,preprint]{aastex}

\slugcomment{Accepted by AJ}

\lefthead{Fornal et~al.}
\righthead{$u'g'r'i'z'$ Photometry of NGC~188}

\begin{document}

\title{A Survey of Open Clusters in the $u'g'r'i'z'$
Filter System: III. Results for the Cluster NGC 188}

\author{
Bartosz Fornal\altaffilmark{1,2,3},
Douglas L. Tucker\altaffilmark{2,4},
J. Allyn Smith\altaffilmark{4,5},
Sahar S. Allam\altaffilmark{2,4,6},
Cristin J. Rider\altaffilmark{7}, and
Hwankyung Sung\altaffilmark{8}
}

\altaffiltext{1}{Institute of Physics,
		     Jagiellonian University,
                 ul. Reymonta 4, 30-059 Krak\'{o}w, Poland
                 }
\altaffiltext{2}{Fermi National Accelerator Laboratory,
                 P.O. Box 500, Batavia, IL 60510
                 }
\altaffiltext{3}{Fermi National Accelerator Laboratory Internship
                 for Undergraduate Physics Majors program}

\altaffiltext{4}{Visiting Observer, United States Naval Observatory, Flagstaff Station, AZ}
\altaffiltext{5}{Department of Physics \& Astronomy, Austin Peay
                 State University, P.O. Box 4608, Clarksville, TN 37044}
\altaffiltext{6}{University of Wyoming,
                 Department of Physics \& Astronomy,
                 Laramie, WY 82071}
                 \label{Wyoming}
\altaffiltext{7}{Department of Physics \& Astronomy, Johns Hopkins University, 
                 3400 N. Charles St., Baltimore, MD 21218-2686
                 }
\altaffiltext{8}{Department of Astronomy \& Space Science, Sejong University, Seoul, 143-747, Korea}

\begin{abstract}
We  continue our series of papers describing the results of a photometric
survey of open star clusters, primarily in the southern hemisphere, taken
in the $u'g'r'i'z'$ filter system. The entire observed sample covered more
than 100 clusters, but here we present data only on NGC 188, which is 
one of the oldest open clusters known in the Milky Way.
We fit the Padova theoretical isochrones to our data.  
Assuming a solar metallicity for NGC~188, we find a distance of $1700\pm100$ pc, 
an age of $7.5\pm0.7$ Gyr, and a reddening $E(B-V)$ of $0.025\pm0.005$.
This yields a distance modulus of $11.23\pm0.14$. 
\end{abstract}


\section{Introduction}

The study of open clusters is extremely important for improving our 
understanding of stellar  and  galactic evolution.
In particular, knowledge of accurate fundamental parameters of star
clusters (i.e., age, distance, reddening, metallicity) is essential 
for many astrophysical calibrations. It also contributes to a better 
explanation of the present Galactic disk properties and its past history.
For example, precise measurements of age and distance of open clusters from the Sun
tell us about their spatial and age distributions, which give
invaluable insight into the disk structure. From these data it is 
possible to infer many characteristics of the Milky Way, like
the thin-disk scale height, horizontal scale length, or the
displacement of the Sun above the Galactic plane (see, \citealt{Piskunov06} 
and \citealt{Bonatto06}). 
Furthermore, data on reddening for open clusters show important
features of the interstellar extinction
(\citealt{Piskunov06}), which provides information on the Milky Way 
gas and dust distribution.
Correctly calculated open cluster parameters are therefore 
essential to constrain Galactic theoretical models.
Among all open clusters, the oldest ones are of special interest.
Their age can be established quite easily and they may be detected at large distances
thanks to their brightest members - strong-lined red giants (\citealt{Friel95}). 
Besides being fine tracers of the present structure and chemistry 
of the Galactic disk, they enable us to look at disk properties at epochs
just after its formation, thus providing a reliable way to probe even the early disk evolution.
Accurately estimated old open cluster basic parameters are then very useful for 
examining the disk-halo connection (\citealt{Janes94}). 
As a result of being such a powerful tool for testing theories of star formation and
metal enrichment in the Milky Way disk, open clusters, especially the
old ones, have been the subject  of intense  studies  for the past several  
decades (besides the papers mentioned above, see,
\citealt{ATTM79,GS84,ATTS94,Bruntt99,WOCS1}).

We undertook a survey  of (mostly)  southern  hemisphere star
clusters  using the    $u'g'r'i'z'$   filter  system (\citealt{Smith02};
\citealt{Rider04}; \citealt{Rodgers06}).
The  initial  effort  in    this  survey delivered   observations   for
approximately 105 open  clusters  and  a few (less than 10)   ``low-density''
globular clusters. 
The    original
motivation  of  the project was  to use  these clusters, which  span a
range  of ages  and metallicities,  to  ``back  calibrate'' the  Sloan
Digital Sky Survey (SDSS) 
(see, \citealt{york00} and \citealt{Adelman06} for a description of the
SDSS; \citealt{Gunn06} for a description of the SDSS 2.5 m telescope;
\citealt{GCRS98} for a description of the SDSS imaging camera; and 
\citealt{fuk96}, \citealt{Hogg01}, \citealt{Ivezic04}, and \citealt{Tucker06}
for a description of the photometric calibration of the SDSS.)
We are now using these data to verify the recent age and
metallicity models presented in \citet{Girardi04} and the prior work of
\citet{Lenz} and to verify and expand upon the $u'g'r'i'z'$ to $UBVRI$
transformations presented in the SDSS standard star paper \citet{Smith02}.
We have recently supplemented these data with observations of northern hemisphere 
clusters using the United States Naval Observatory 1.0 m telescope.

In this, the third paper  of our series, we present our results  for the open cluster
NGC~188. We chose this particular cluster because it is one of the oldest open 
clusters known in our galaxy. As described in \citet{Bonatto05}, it is located 
quite far from the galactic disk and contains a couple of hundred member stars. Its 
field is not heavily contaminated by background stars and it is also relatively 
free from dust \citep{Bonatto05}.
All this makes NGC 188 a perfect target for testing stellar evolution models.

NGC 188 was discovered in 1831 by John Herschel. Since then, Ivan King 
noticed in 1948 that this cluster is very interesting since its angular diameter is large when
compared to the faint apparent magnitudes of its brightest stars (see,
\citealt{Sandage62}). In 1956 Sidney 
van den Bergh obtained the first results suggesting that NGC 188 
belongs to the group of the oldest clusters (see, \citealt{Sandage62}).

NGC 188 has been very carefully studied throughout the years, but the derivation of its 
characteristic parameters has had an erratic history. Table \ref{table} presents a summary of
past results on NGC~188. This table includes the 
reference paper (col. [1]), estimated distance modulus, in the brackets
the corresponding distance (corrected for the reddening $E(B-V)$) (col.  [2]), age (col
[3]), $E(B-V)$ (col. [4]), metallicity $[Fe/H]$ (col. [5]), the technique used (col. [6]), 
and finally additional comments at the end.

\placetable{table}

One of the first studies of this cluster to determine age, distance and reddening is found 
in \citet{Sandage62} and \citet{Sandage_2_62}. 
In these two papers, Sandage estimated the age of this cluster to be $14-16$ Gyr (based on 
\citealt{Hoyle59} stellar models), the distance modulus $m-M$ to be 10.95, and the reddening 
$E(B-V)$ to be between 0.03 and 0.07, all these values depending on the chemical composition.

Revised values of NGC 188 parameters based on $UBV$ photometry are presented in \citet{Eggen69}, 
who give an age of 10 Gyr, a distance modulus of 10.85, a reddening $E(B-V)$ of 0.09, and a metallicity $[Fe/H]$ of 0.07.
During the following two decades estimates of the age varied from 3.6 Gyr (\citealt{Torres71}) 
and 4.3 Gyr (\citealt{Twarog78})
up to 10 Gyr (\citealt{VandenBerg85}, \citealt{Adler85}). Estimates of the metallicity 
ranged from $[Fe/H]=-0.6$ (\citealt{Jennens75}) to as much as $[Fe/H]=0.6$ (\citealt{Spinrad70}). 
The estimates for the distance modulus fell within the range of 10.8 (\citealt{Pat78}) to 
12.0 (\citealt{Jennens75}), while estimates for the reddening $E(B-V)$ varied from 0.04 (\citealt{Jennens75}) 
to 0.15 (\citealt{Spinrad70}).

The analysis of color-magnitude diagrams performed by \citet{Twarog89} yielded an age between 6 
and 7 Gyr, a distance modulus of 11.50 and a high $E(B-V)$ value of 0.12.
Besides the \citet{Hobbs90} value of $E(B-V)$ = 0.10 no other estimation of the reddening was made 
until \citet{Carraro94} paper which yielded values 0.03 and 0.04, depending on the adopted stellar evolution model. 
The value of age varied from 6 Gyr (\citealt{Paez90}, \citealt{Dinescu95}) to 7.7 Gyr (\citealt{Hobbs90}). Estimations
of distance modulus ranged from just a little over 11.0 (\citealt{Branly96}) up to 11.5 (\citealt{Dinescu95}). 
The estimated metallicity was always close to solar.

A more recent, high-precision $UBVRI$ CCD photometry study by \citet{Sarajedini99} provided a 
color-magnitude diagram which 
extends almost from the red giant branch tip to approximately 5 mag below the main sequence 
turn-off. The final conclusion was that there is a considerable offset between the photometric zero point
of these results and those from \citet{Eggen69}, whose photometric scale was used in
all previous photometric studies of NGC 188. Reddening $E(B-V)$ was found to be equal to
$0.09\pm0.02$ and the distance modulus to be $11.44\pm0.08$. The metallicity was 
assumed to be solar, based on the result $[Fe/H] = -0.04\pm0.05$ from \citet{vonHippel98}.
The fitted isochrones yielded an age of 
$7.0\pm0.5$ Gyr. Moreover, the data indicate that there exists a mass segregation with the most 
massive stars \((M/M_\odot > 1.1)\) more centrally concentrated comparing to the least 
massive ones ($0.8 > M/M_\sun > 0.65$).

A metallicity for NGC~188 close to solar was verified in \citet{Friel02}, \citet{Randich03}, and \citet{Worthey03}.
The values of other parameters were revised in three of the most recent papers --- \citet{VandenBerg04}, 
\citet{Bonatto05}, and \citet{Haroon04} --- by
fitting theoretical isochrones to the color-magnitude diagram.
In the first one, after adopting a reddening $E(B-V)$ of 0.087 from \citet{Schlegel98} dust maps, \citet{VandenBerg04} 
determined an age of 6.8 Gyr and 
a distance modulus of 11.40 (distance 1685 pc). In the second of these papers, \citet{Bonatto05} found a slightly higher
age of 7.1 Gyr and a smaller distance modulus of 11.10; their best-fit isochrone
indicates $E(B-V) = 0.00$ (which gives a distance of 1660 pc). Finally, both of these age values 
fall within the range stated in the third paper, \citet{Haroon04}. The
results therein indicate an age between 6 and 8 Gyr, more likely close to 8 Gyr. The 
fitted isochrone yields a distance modulus of $11.26\pm0.05$.
An abnormally high value of reddening, $E(B-V)=0.17$, lowers the resulting distance down 
to only $1415\pm35$ pc.

The most recent star catalog for NGC 188 is presented in \citet{Stetson04}.
It contains detailed information on more than 9000 stars in the field of the cluster 
based on all available studies. In addition,  half of those stars have revised photometry.

Ernst Paunzen and Jean-Claude Mermilliod's {\tt  webda} online open  cluster database,\footnote{{\tt
http://www.univie.ac.at/webda/webda.html}} which provides a compilation of data from several sources,
currently (August 2006) lists the age of NGC 188 as 4.3 Gyr with  a
distance of 2047 pc, a distance modulus of 11.81, a reddening of $E(B-V)=0.082$, and
a metallicity of $[Fe/H] = -0.02$.

In  the following sections of this paper we present details of the instrumentation
and  observations  (\S2),   data reduction  and analysis  techniques
(\S3), isochrones (\S4), results (\S5), and a summary (\S6).

\section{Instrumentation and Observations}

\subsection{$u'g'r'i'z'$ Filter System}

The five filters of the $u'g'r'i'z'$ system have effective wavelengths 
of 3540\AA, 4750\AA,  6222\AA, 7632\AA, and  9049\AA , respectively, at
1.2 airmasses.\footnote{Note that the  $g'$ filter has been determined
to have  an effective wavelength 20  \AA\, bluer  than that originally
quoted by \citet{fuk96}.}   They cover the  entire wavelength range of
the  combined   atmosphere and CCD  response. Their  construction  is
described  in   \citet{fuk96}. 
The most important characteristics of the $u'g'r'i'z'$ filters and 
the $u'g'r'i'z'$ magnitude system are outlined in \citet{Rider04}. 
For a more detailed explanation they refer to \citet{OG83}, \citet{fuk96},
and \citet{Smith02}.
The   $u'g'r'i'z'$ standard star network consists  of  158 stars
distributed  primarily along the  celestial  equator and the  northern
celestial    hemisphere \citet{Smith02}.
Efforts are in place and nearing completion for extending this
network both fainter and redder \citep{Smith07} and into the 
southern hemisphere \citep{Smith06}.

\subsection{ Telescope and Observations }

For this paper, we are using data from the USNO 1.0 m telescope. 
The observations for NGC~188 were obtained on 2002 November 5 (UT)
as part of a four-night observing run. 
An overview of the observing circumstances is
given in Table~\ref{obs}.

\placetable{obs}

All of the observations were direct exposures with a thinned, UV-AR coated, 
Tektronix TK1024 CCD operating at a gain of 7.43 $\pm$ 0.41 \(e^{-}\) ADU\(^{-1} \)
with a read noise of 6.0 \(e^{-}\). This CCD is similar to the CCDs used in 
the SDSS 2.5 m telescope's imaging camera and the CCD used by the 0.5 m
SDSS Photometric Telescope at Apache Point Observatory.
This is the detector that defines the $u'g'r'i'z'$ standard star 
network. The camera scale of 0.''68 pixel\(^{-1} \) for this 1024 x 1024 detector 
produces a field of view of 11.54 arcmin.

During a typical night at the telescope, we generally observed four to
five standard fields at the start of the night to determine the 
extinction. Following this, we would usually observe two to three
target fields and then alternate between two to three standard 
and target fields through the remainder of the night, finishing 
with a longer run of standards (usually four to five). In 
general, two or three standard fields were observed several times 
each night to monitor extinction manually at the telescope and to 
look for changes in the photometricity of the sky. These values
were compared with the ``all-sky'' extinction values determined
later during reduction process. Additional fields were observed 
throughout the night, near the meridian and at high air mass, in 
order to provide a good color spread to solve for instrumental 
color terms.

\section{Data Reduction}

We performed reductions using version  {\tt v8.3} of the SDSS software
pipeline {\tt  mtpipe} (see, \citealt{Tucker06}).  This software processes
the images and performs aperture photometry.  It also determines
photometric zero points based on observations of standard star target
fields (i.e., stars from \citealt{Smith02}) and applies the fitted 
photometric equations to the aperture photometry lists.  A detailed description of 
{\tt mtpipe} and specific details of how it is used for the
open cluster survey can be found in the first paper of this
series (\citealt{Rider04}).  
In  general, we follow the reduction procedures  outlined by \citet{Rider04} 
with the following exception --- we use an aperture with 
radius of $7.43''$.
This aperture size was chosen because it was used in the calibration of the
southern $u'g'r'i'z'$ standard star network (\citealt{Smith03}, \citealt{Smith06}), 
and it produced good
fits to the photometric equations for the data from the USNO telescope. 

The night characterization  data  for each  of the  photometric nights
included in this study are given in Table~\ref{nightChar}.
These data include the filter (col. [1]), zero   
points  (col. [2]), and the first-order extinction terms (col. [3]).
Columns [4] and [5] give the rms errors for, and the number of, 
the standard stars observed that night which were used in the photometric solutions.   
In a footnote, we also list the second-order extinction terms  
derived in \citet{Smith02}.

\placetable{nightChar}

After running all the data through {\tt
mtpipe},  we had four lists of calibrated  $u'g'r'i'z'$  photometry
for NGC~188 from the USNO 1.0 m telescope. 
Two of these lists were excluded from further analysis after comparing 
zeropoint offsets
and finding problems in the photometry in at least one filter for 
these two lists.
We combined the contents 
of the two remaining lists together into a single list. 
We did  the combining by  assigning  each star  in   the
combined list the ``best'' magnitude of its corresponding entries from
both lists.  In this context, ``best'' refers to
the magnitude that has  the smallest photon  noise without a saturation 
flag being set. This was done on a
filter-by-filter basis, so a star's  best $u'$ magnitude and best $r'$
magnitude may come from  different lists.  (Star entries were  matched
by position between  lists using a $\pm$2~arcsec box  in RA  and DEC.)
Magnitude entries which had been  flagged by {\tt mtpipe} as  being
saturated or which had  poorly determined values (magnitudes $<0$  or
$>100$) were excluded from the combine procedure.

Since the {\tt webda} value of NGC~188's angular diameter is 17.0 arcmin, 
only stars within a cluster radius of 8.5~arcmin were included  
in the final list in order to reduce field-star contamination.
However, the limiting diameter for NGC~188 appears to be almost three times
larger than its {\tt webda} value (see, \citealt{Bonatto05}).
The same paper shows (see, Fig. 1 of \citealt{Bonatto05}) that  
field-star contamination in the region inside a radius $R=8.5$ arcmin 
cannot be neglected.
On the other hand, this field is characterized by a high contrast 
between cluster and background star density (see, Fig. 2 of \citealt{Bonatto05}).
Anyhow, our analysis, which deals primarily with NGC~188's fundamental 
parameters, is not influenced by the facts mentioned above. 
Where possible, we assigned cluster membership probabilities from 
\citet{Platais03}; these membership probabilities were  
based on proper motions given in the same paper. 
In our analysis, we assumed stars with \citet{Platais03} membership
probabilities over 50\% to be cluster members and those with membership 
probabilities less than 50\% to be non-members.

Table \ref{dataUSNO} provides the available data from the 
USNO 1.0 m telescope for NGC 188. 
This table includes our internal ID number for each
star (col. [1]), the {\tt  webda} star ID number (col.  [2]), RA (col
[3]), DEC (col. [4]), $ u'g'r'i'z'$ magnitudes (cols. [5], [6], [7],
[8], \& [9], respectively), $ u'g'r'i'z'$ magnitude (photon noise)
errors (cols. [10], [11], [12], [13], \& [14], respectively),
$u'g'r'i'z'$ saturation flags (cols.  [15], [16], [17], [18], \& [19],
respectively), the cluster membership probability from \citet{Platais03}
(col. [20]), and comments (col. [21]). Since our data on NGC~188
come only from one telescope, our internal numbering scheme (col. [1])
is just a running ID for the entries in Table \ref{dataUSNO}.

\placetable{dataUSNO}

\section{Isochrones}
 
We fit the theoretical SDSS isochrones
and   metallicity  curves     from   \citet{Girardi04} to our observational data of NGC~188,
from where we derive the fundamental parameters of this cluster.
The  input physics  for
these models are based upon a \citet{Mihalas90} equation of state
at temperatures $T < 10^{7}$\ K and a fully-ionized gas equation of state
at  higher temperatures; electron  screening   is incorporated in  the
reaction rates.   The  theoretical evolutionary tracks  were converted
into  the  SDSS photometric  system  using   the SDSS 2.5~m  telescope
$ugriz$       filter        response     functions     and     the
no-overshoot  ATLAS9  synthetic   atmospheres of   \citet{castelli97}.

Note that the \citet{Girardi04} SDSS isochrones were created using the
SDSS 2.5~m telescope  $ugriz$ filter  system, which differs   slightly
from the USNO~1.0~m   $u'g'r'i'z'$  filter system  (see,  for  example
\citealt{abazajian03}); furthermore, \citet{Girardi04} assume that the
SDSS 2.5~m $ugriz$   system is a    perfect $AB$ system. In  order  to
compare  the  isochrones to our   data   we first  had to  adjust  the
isochrones for the known  deviation of the SDSS  2.5~m telescope from a
true  $AB$ system.  This was  done  using the following equations  (D.
Eisenstein, private communication):
\begin{eqnarray}
u(AB,2.5~m) & = & u(2.5~m) - 0.040\;, \\
g(AB,2.5~m) & = & g(2.5~m) - 0.009\;, \\
r(AB,2.5~m) & = & r(2.5~m)\;, \\
i(AB,2.5~m) & = & i(2.5~m) + 0.017 \;, \\
z(AB,2.5~m) & = & z(2.5~m) + 0.035\ .
\end{eqnarray}
(The values of the $AB$ offsets in these equations are preliminary and
future refinement at the $\pm$ 0.01--0.02~mag level are possible.)

Next, the isochrones were converted from the SDSS 2.5 m $ugriz$ system
into the USNO 1.0~m $u'g'r'i'z'$ system by making use of the following
relations \citep{Tucker06} :
\begin{eqnarray}
u(2.5~m) & = & u'\;, \\
g(2.5~m) & = & g' + 0.060((g'-r')-0.53)\;, \\
r(2.5~m) & = & r' + 0.035((r'-i)'-0.21)\;, \\
i(2.5~m) & = & i' + 0.041((r'-i')-0.21)\;, \\
z(2.5~m) & = & z' - 0.030((i'-z')-0.09)\ .
\end{eqnarray}

The stars were then dereddened with the following equations, which make use of the 
extinction coefficients given by \citet{Girardi04} for $A_{V}=1.0$, $R_{V}=3.1$, 
and $T_{eff}=5777$\ K:
\begin{eqnarray}
   u' & = & u'_{red} - 4.879 \times E(B-V)\;, \\
   g' & = & g'_{red} - 3.708 \times E(B-V)\;, \\
   r' & = & r'_{red} - 2.722 \times E(B-V)\;, \\
   i' & = & i'_{red} - 2.089 \times E(B-V)\;, \\
   z' & = & z'_{red} - 1.519 \times E(B-V)\ .
\end{eqnarray}

\section{Results}

\begin{figure}
\scalebox{0.8}{\includegraphics{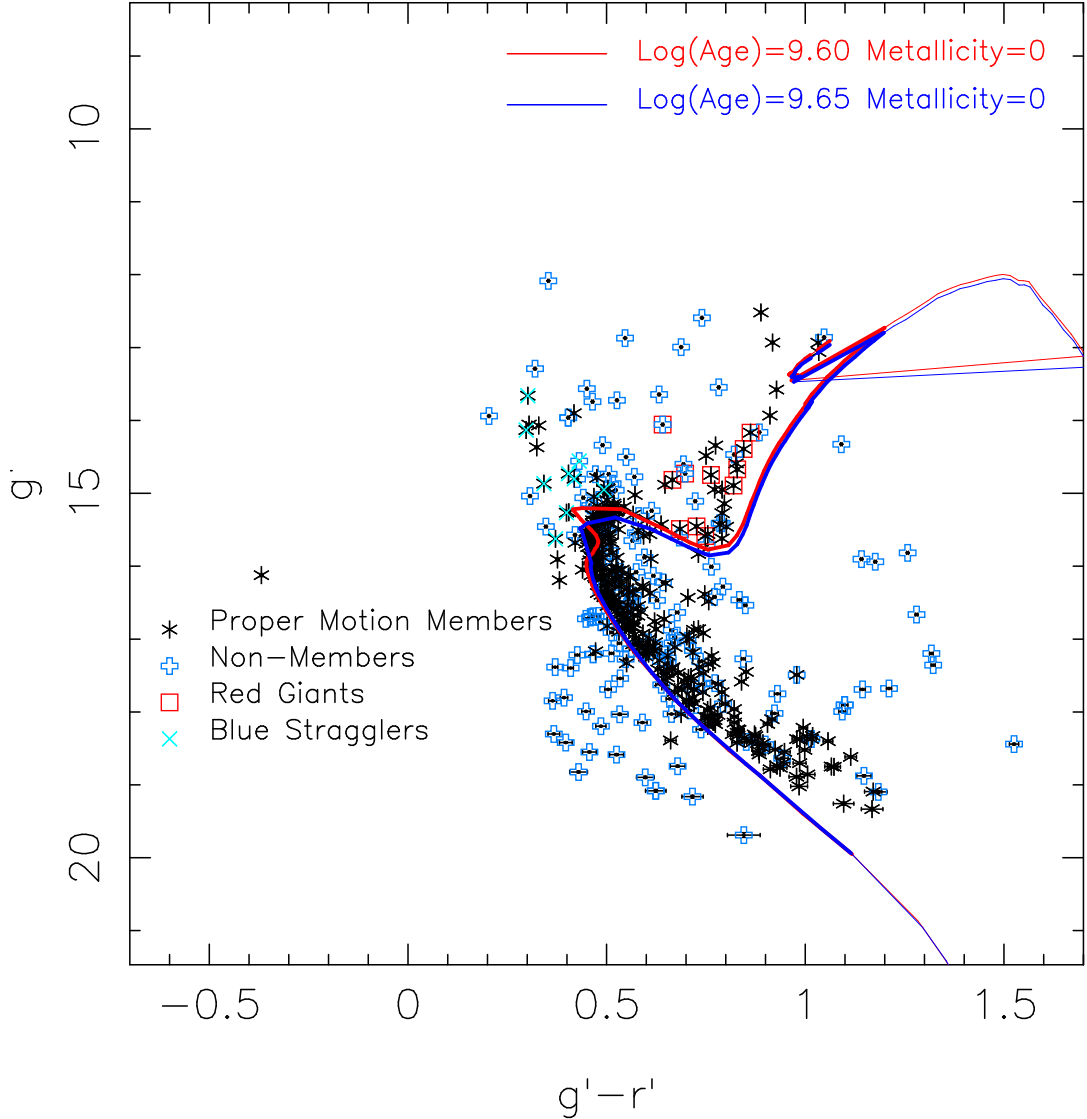}}
\caption{The observed (not dereddened) $g'$ vs. $(g'-r')$ color-magnitude diagram for NGC~188 
using data from the USNO 1.0 m telescope. The solid lines are the 4.0 
and 4.5~Gyr isochrones (distance 2047 pc, $E(B-V)=0.082$, $[Fe/H]=0.0$) from  
\citet{Girardi04} which bracket the  {\tt webda} value of 4.3~Gyr.  
Stars identified as cluster members based upon their \citet{Platais03} proper
motion membership probabilities (probability $>$50\%) are indicated by asterisks;
those identified as non-members (probability $<$50\%), by thick plus signs. Those
stars classified in {\tt webda} as red giants or as blue stragglers are marked by
square symbols or by $\times$ symbols, respectively.
}
\label{webda_fit}
\end{figure}

\begin{figure}
\scalebox{0.8}{\includegraphics{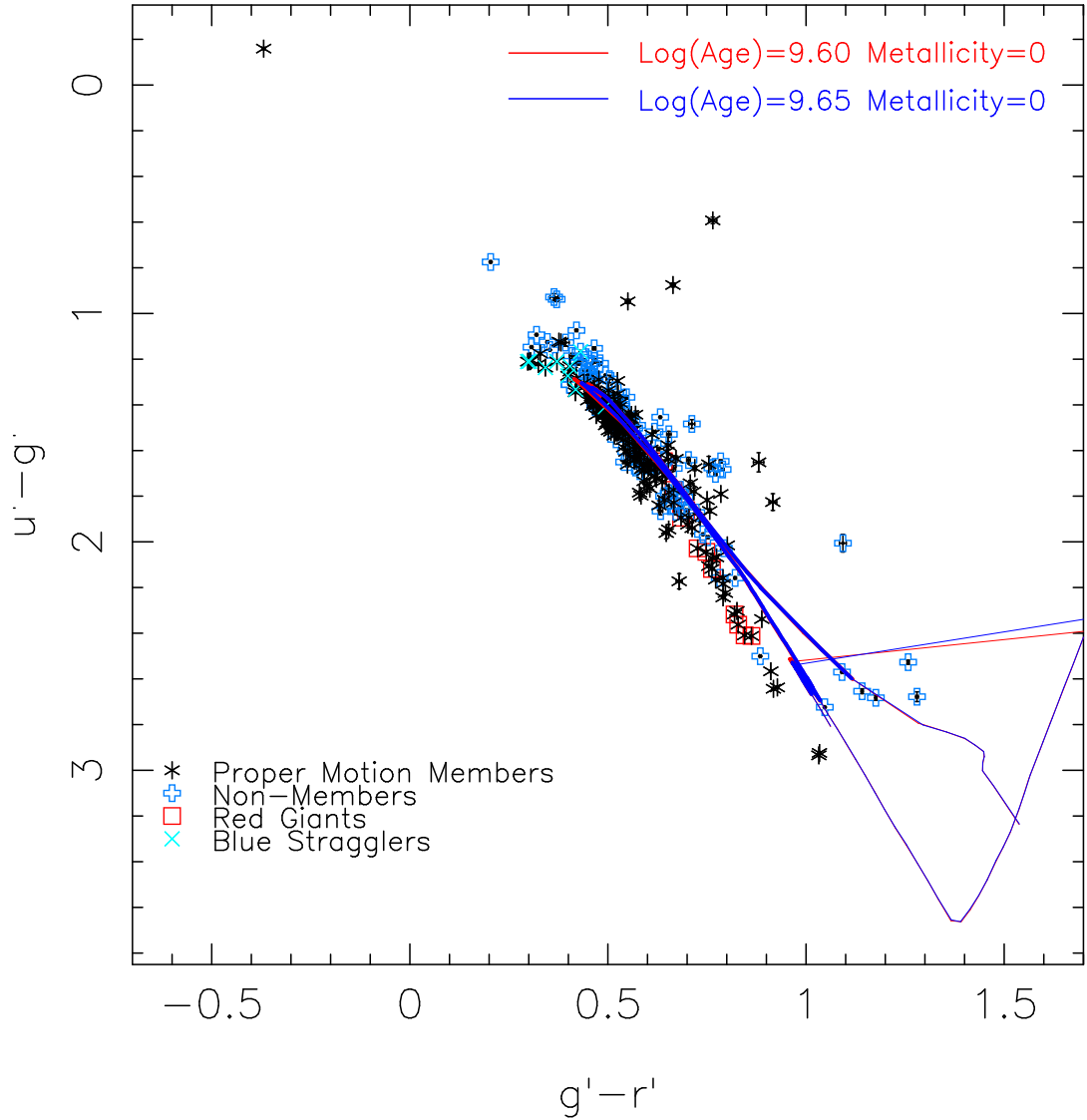}}
\caption{The observed (not dereddened) $(u'-g')$ vs. $(g'-r')$ color-color diagram 
for NGC~188 using data from the USNO 1.0 m telescope. 
The solid lines are the 4.0 
and 4.5~Gyr isochrones (distance 2047 pc, $E(B-V)=0.082$, $[Fe/H]=0.0$) from  
\citet{Girardi04} which bracket the  
{\tt webda} value of 4.3~Gyr. 
The symbols are the same as in Fig. \ref{webda_fit}.
}
\label{webda_fit_cc}
\end{figure}

We initially adopted the current (August 2006)
{\tt webda} values for distance, age, reddening $E(B-V)$, and metallicity $[Fe/H]$ 
as a first guess for these parameters for NGC~188. Figure \ref{webda_fit}
shows the color-magnitude diagram for our set of data points overplotted with the
theoretical isochrones which bracket the {\tt webda} value of age for NGC 188.
Figure \ref{webda_fit_cc} shows the same comparison on the color-color diagram.

\placefigure{webda_fit}

\placefigure{webda_fit_cc}

These isochrones do not fit 
our data. There is no match, especially for the red giant branch.
We therefore attempted to find the best fitting theoretical isochrone by 
fitting the data by eye. 
We decided that the best way of doing this would be to
reduce as much as possible the number of unknown cluster parameters.
The {\tt webda} states NGC~188's metallicity as $[Fe/H]=-0.02$. All the latest papers
also indicate that its value is close to zero: $-0.02$ (\citealt{Twarog97}), 
$-0.04$ (\citealt{vonHippel98}), $-0.10$ (\citealt{Friel02}), 
0.00 (\citealt{Randich03}), 0.075 (\citealt{Worthey03}).
Apart from that, we noticed that change in metallicity $[Fe/H]$ of the order of 0.4 dex doesn't 
influence the shape of the Girardi isochrone by large amount.
We therefore assumed a solar metallicity for NGC~188.
We were then left with three free parameters. Since a change in distance shifts
the isochrone on the color-magnitude diagram only up and down, we would be 
able to fit the isochrone easily after determining the value of age 
or reddening $E(B-V)$ in an independent way. We decided to do this for the reddening.

\begin{figure}
\scalebox{0.8}{\includegraphics{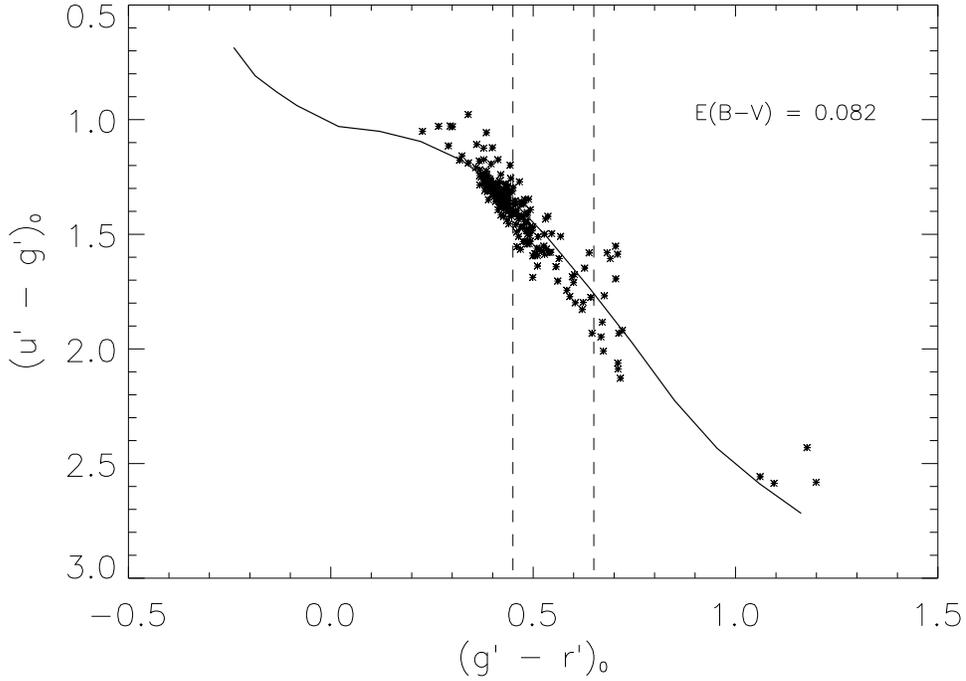}}
\caption{The dereddened $(u'-g')_{0}$ vs. $(g'-r')_{0}$ color-color diagram 
(assuming {\tt webda} value of 
$E(B-V)=0.082$) for NGC~188's main sequence stars 
using data from the USNO 1.0 m telescope. The solid curve is the 
theoretical Zero Age Main Sequence curve obtained after applying 
the transformation equations from \citet{Karaali05} to the $U-B$ 
and $B-V$ values for the ZAMS taken from Table 3.9 of Binney, J., 
Merrifield, M.``Galactic Astronomy''.
The two vertical dashed lines represent the main sequence band between 
$(g'-r')_{0}=0.45$ and 0.65,
which is the region where the data and ZAMS should fit best.
} 
\label{red_0.082}
\end{figure}

From the color-color diagram we were able to infer the value of reddening $E(B-V)$ by
comparing our data points corrected for the reddening with the theoretical 
predictions for the main sequence 
isochrone on the color-color diagram for our set of filters.
In order to analyze only the main sequence stars and eliminate red giants, we took 
into consideration stars with magnitudes higher than 15 and lower than 17 mag.
We over-plotted these data points with the theoretical curves of the $(g'-r')$ vs. $(u'-g')$ 
dependence for the Zero Age Main Sequence (ZAMS) stars. The $U-B$ and $B-V$ values for 
the ZAMS were taken from Table 3.9 of Binney, J., Merrifield, M. ``Galactic Astronomy''. 
The $g'-r'$ and $u'-g'$ values were calculated using the transformation equations from 
\citet{Karaali05}.
Figure \ref{red_0.082} shows the result obtained for the {\tt webda} value of reddening
$E(B-V)=0.082$.

\placefigure{red_0.082}

\begin{figure}
\scalebox{0.8}{\includegraphics{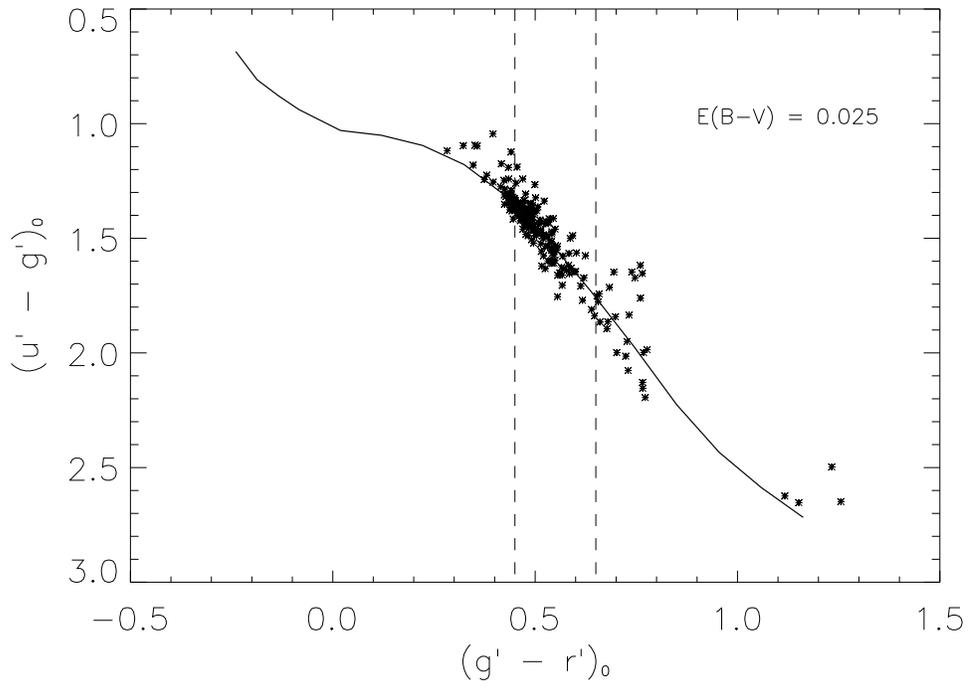}}
\caption{The dereddened $(u'-g')_{0}$ vs. $(g'-r')_{0}$ color-color diagram (assuming 
$E(B-V)=0.025$) for NGC~188's main sequence stars 
using data from the USNO 1.0 m telescope. The solid curve and dashed lines
are as in Figure 3.
} 
\label{red_0.025}
\end{figure}

The theoretical curve seems to match the data points quite well, but this 
was not the best fit. The best fit of the theoretical ZAMS curve to the data points 
was obtained for $E(B-V)=0.025$ (Figure \ref{red_0.025}), which is noticeably 
different from the {\tt webda} value. 

\placefigure{red_0.025}

Having fixed the values of reddening $E(B-V)$ and metallicity $[Fe/H]$, fitting the theoretical 
isochrone to our color-magnitude data points became straightforward. 
First, we chose the age for which the fitted isochrone had the same color of the turn-off
as our data points. Then, by shifting the isochrone up and down (changing
the distance), we finally obtained the best fit. 

\begin{figure}
\scalebox{0.8}{\includegraphics{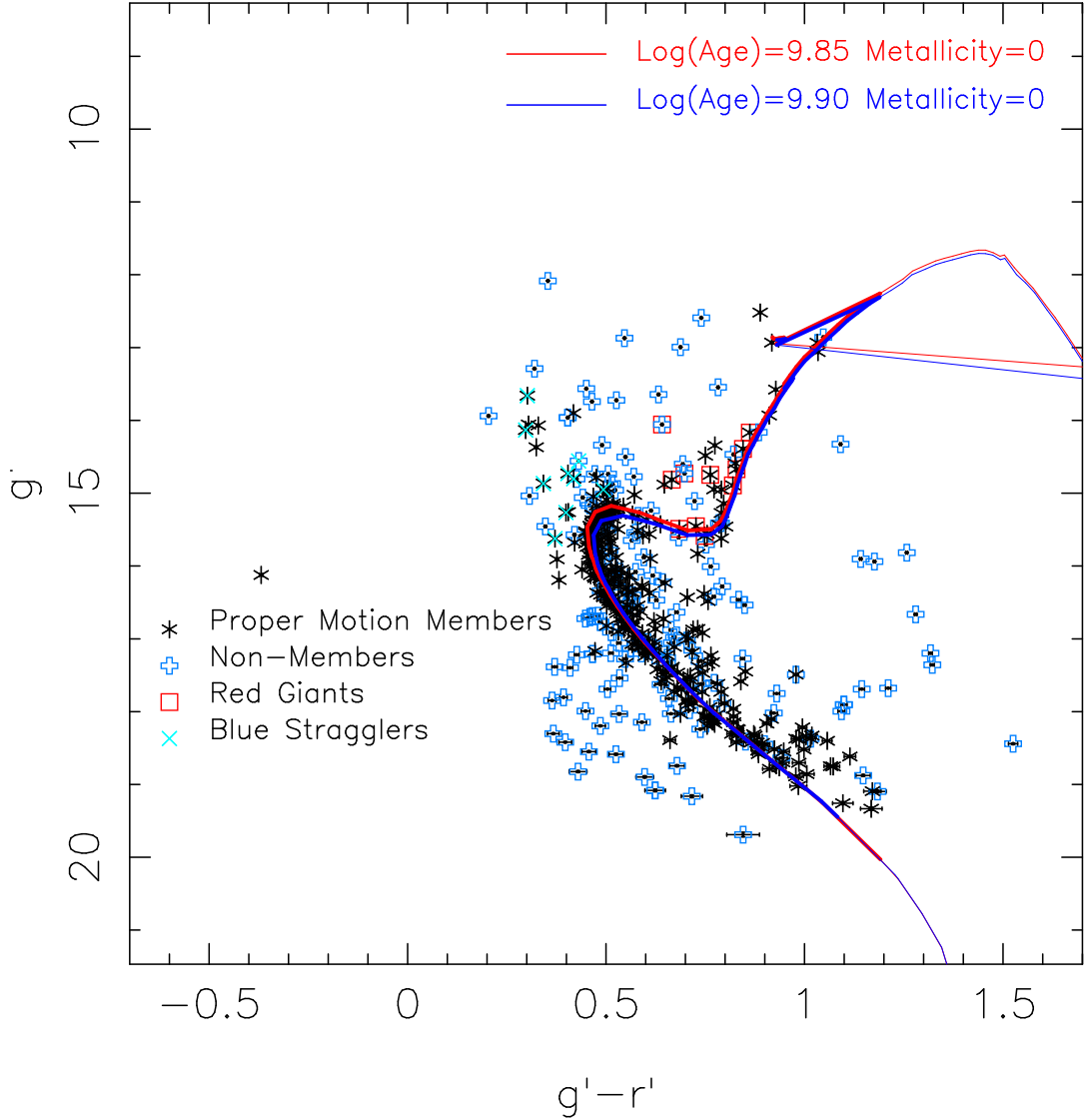}}
\caption{The observed (not dereddened) $g'$ vs. $(g'-r')$ color-magnitude diagram for NGC~188 
using data from the USNO 1.0 m telescope. 
The solid lines are the 7.1 and 7.9~Gyr isochrones (distance 1700 pc, 
$E(B-V)=0.025$, $[Fe/H]=0.0$) from \citet{Girardi04} which bracket our
value of 7.5~Gyr.
The symbols are the same as in Fig. \ref{webda_fit}.
This time a much better isochrone fit is obtained than for the {\tt webda} values.
Even the red giant branch is precisely restored. 
}
\label{my_fit}
\end{figure}

\begin{figure}
\scalebox{0.8}{\includegraphics{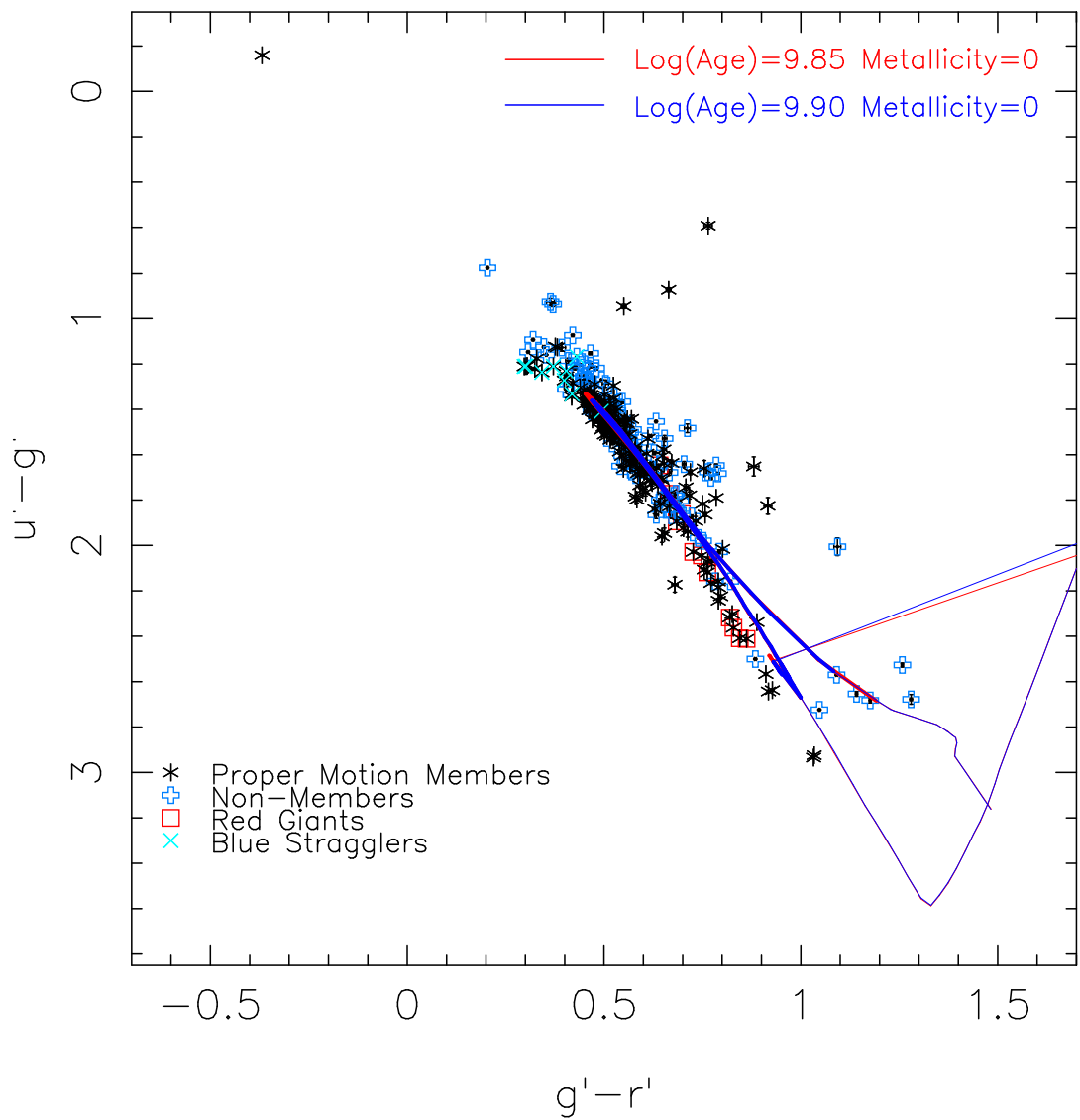}}
\caption{The observed (not dereddened) $(u'-g')$ vs. $(g'-r')$ color-color diagram for NGC~188 
using data from the USNO 1.0 m telescope.
The solid lines are the 7.1 and 7.9~Gyr isochrones (distance 1700 pc, 
$E(B-V)=0.025$, $[Fe/H]=0.0$) from \citet{Girardi04} which bracket our
value of 7.5~Gyr.
The symbols are the same as in Fig. \ref{webda_fit}.
}
\label{my_fit_cc}
\end{figure}

Based upon the  \citet{Girardi04} isochrones we find
that the cluster data  are consistent with a distance of $1700\pm100$ pc,
an age in the range of 6.8 to 8.2 Gyr, and a reddening $E(B-V)$ between 
0.020 and 0.030; however, we find that a distance  of 1700~pc, 
an age of 7.5~Gyr, and a reddening $E(B-V)$ of 0.025 best fit our data
(Figure \ref{my_fit}). These values yield a distance modulus of $11.23\pm0.14$. 
The corresponding isochrone fits on the color-color diagram are shown 
in Figure \ref{my_fit_cc}.

\placefigure{my_fit}

\placefigure{my_fit_cc}

\begin{figure}
\begin{picture}(400,500)
\put (45,0){\includegraphics[angle=0,scale=0.30]{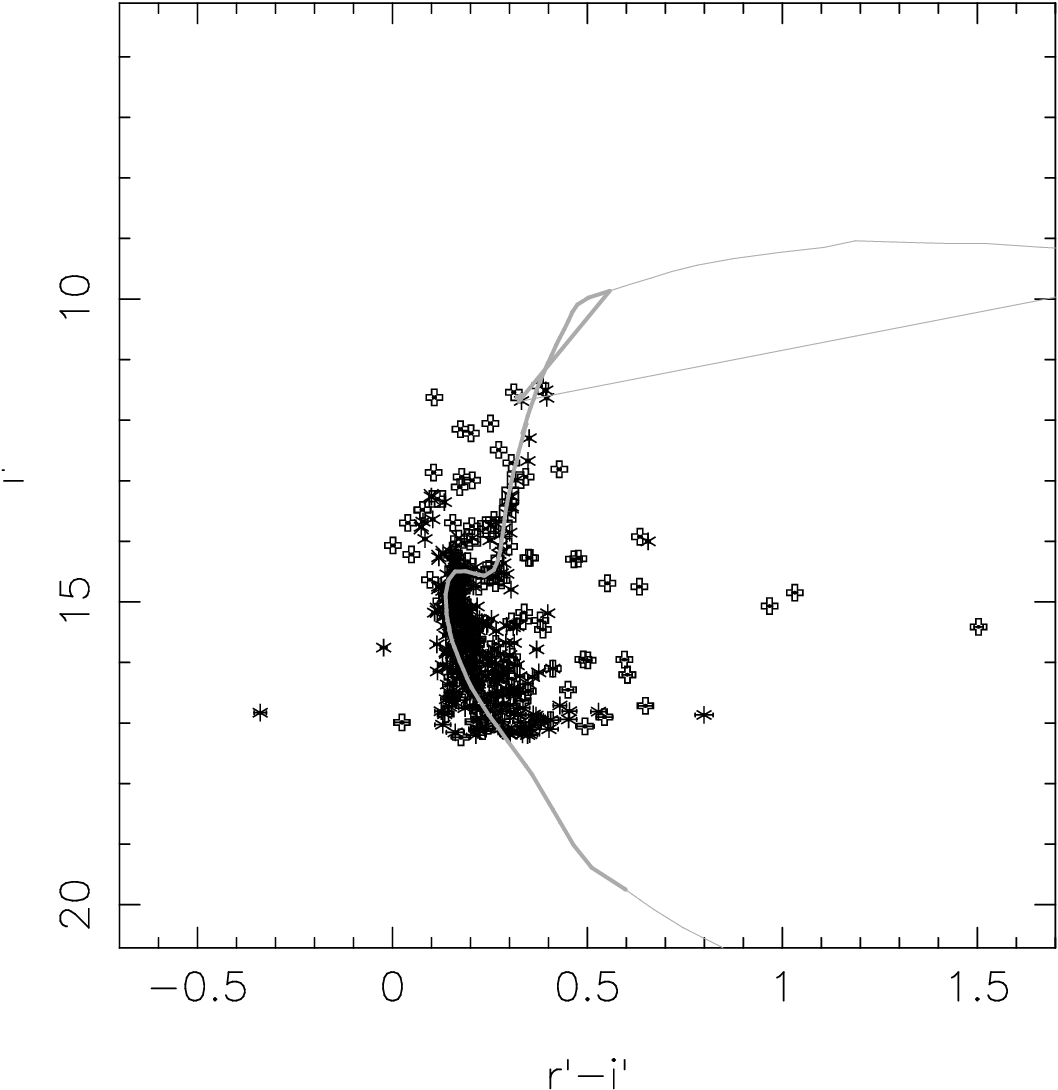}}
\put (250,0){\includegraphics[angle=0,scale=0.30]{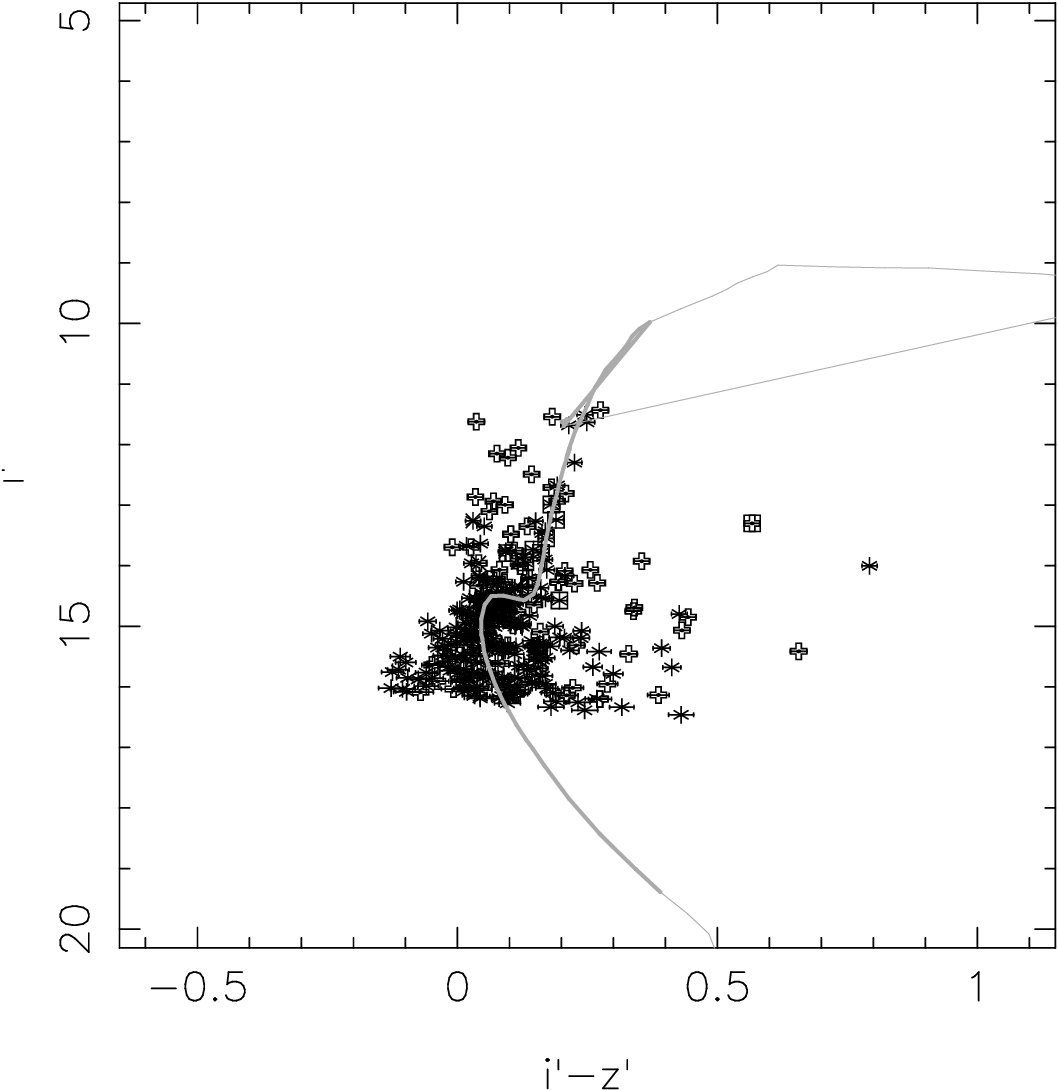}}
\put (45,180){\includegraphics[angle=0,scale=0.30]{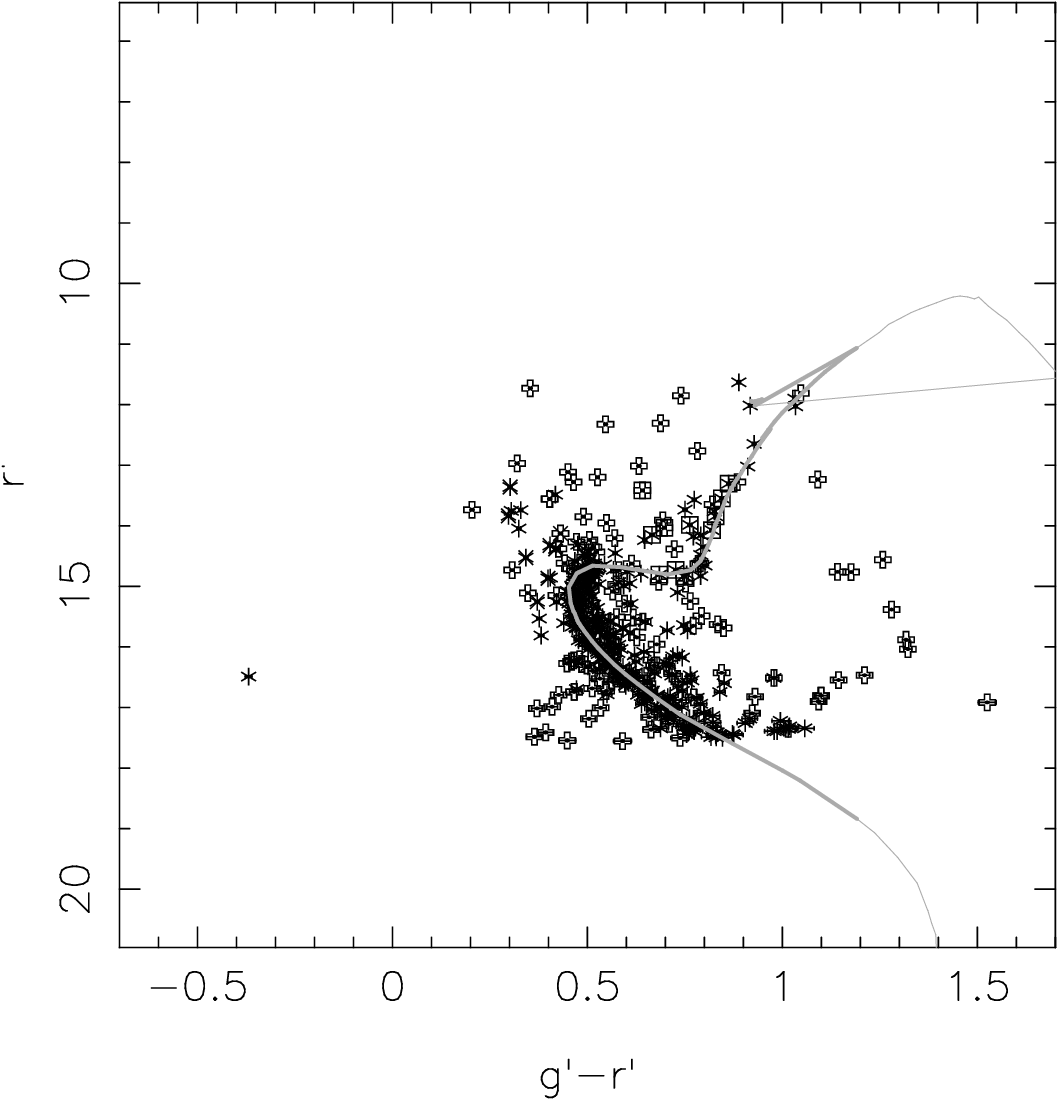}}
\put (250,180){\includegraphics[angle=0,scale=0.30]{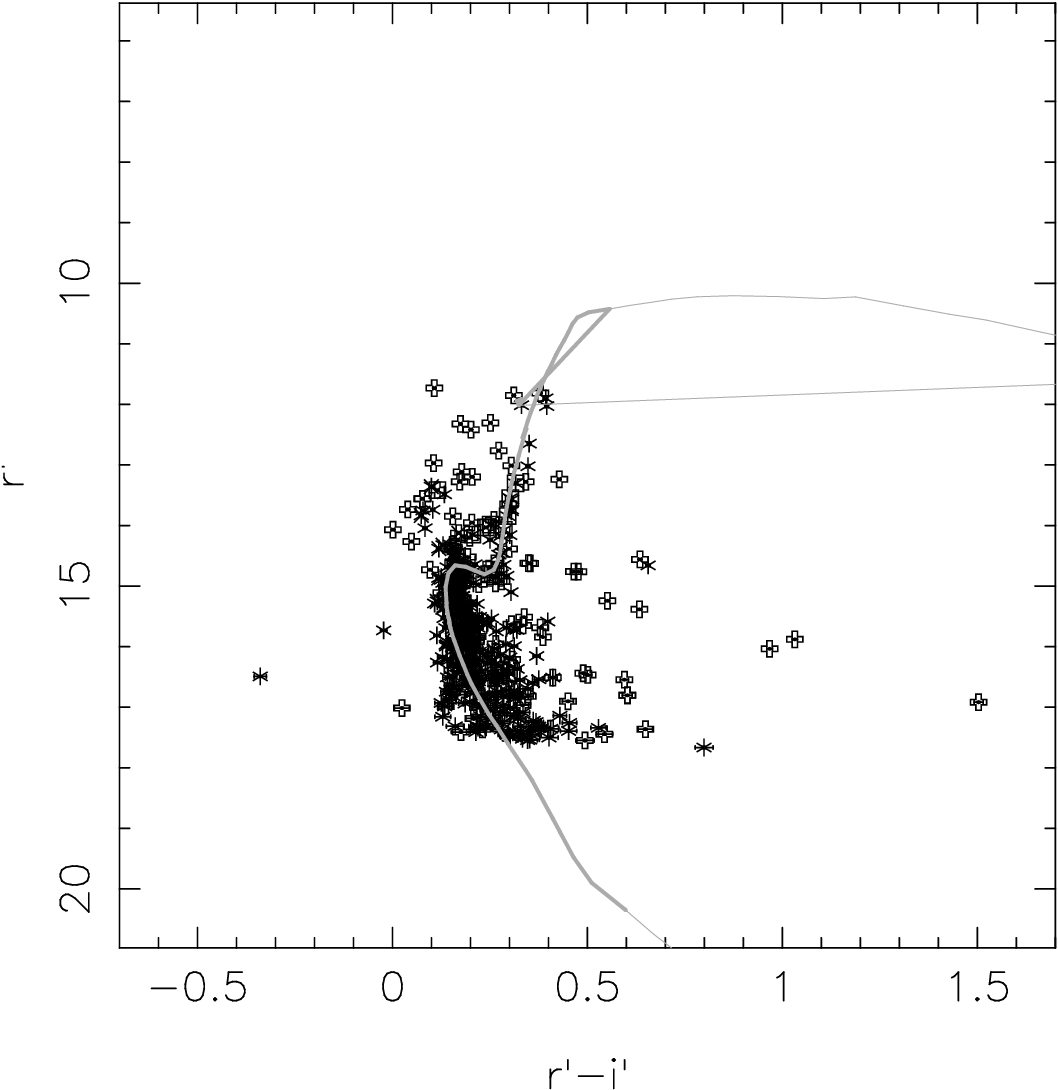}}
\put (45,360){\includegraphics[angle=0,scale=0.30]{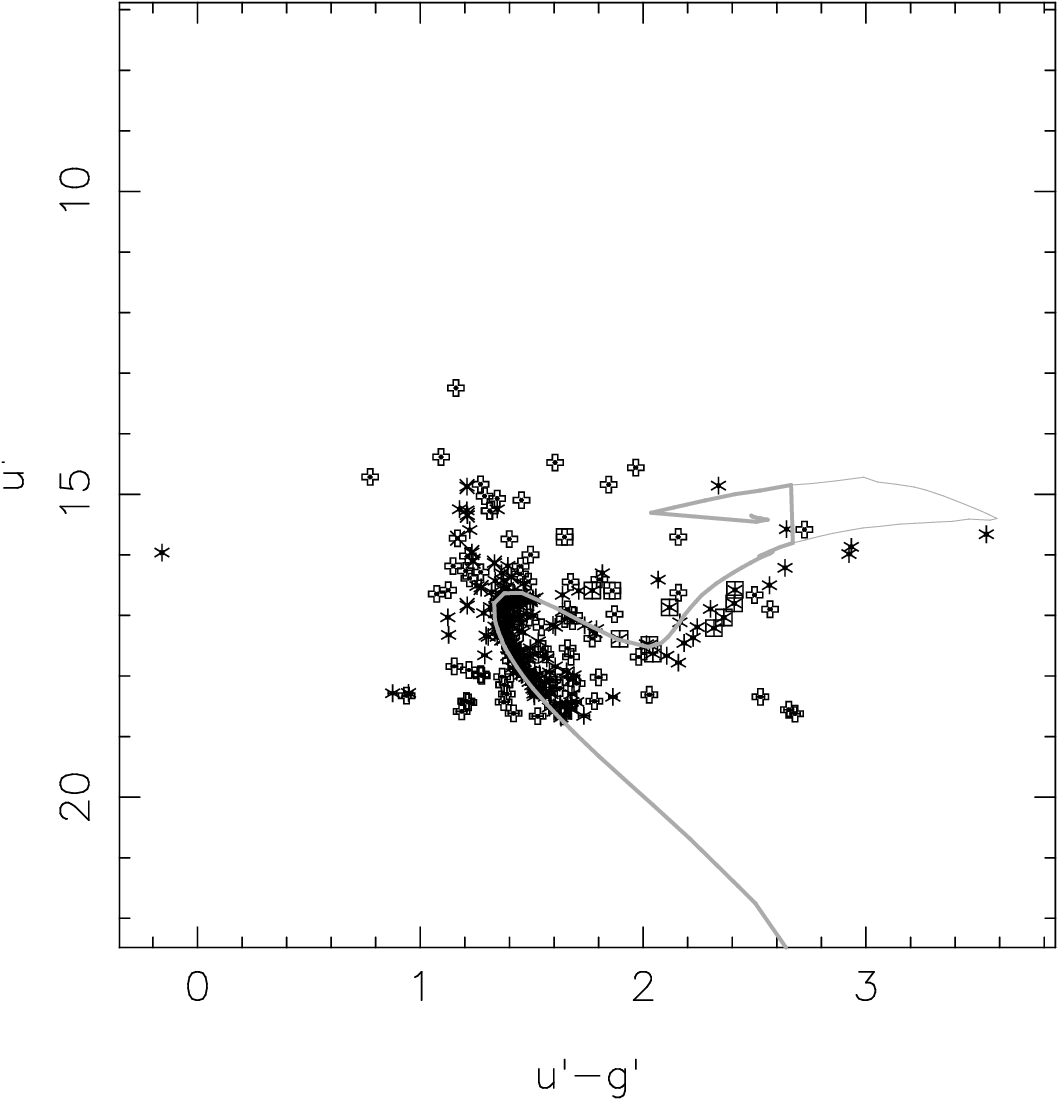}}
\put (250,360){\includegraphics[angle=0,scale=0.30]{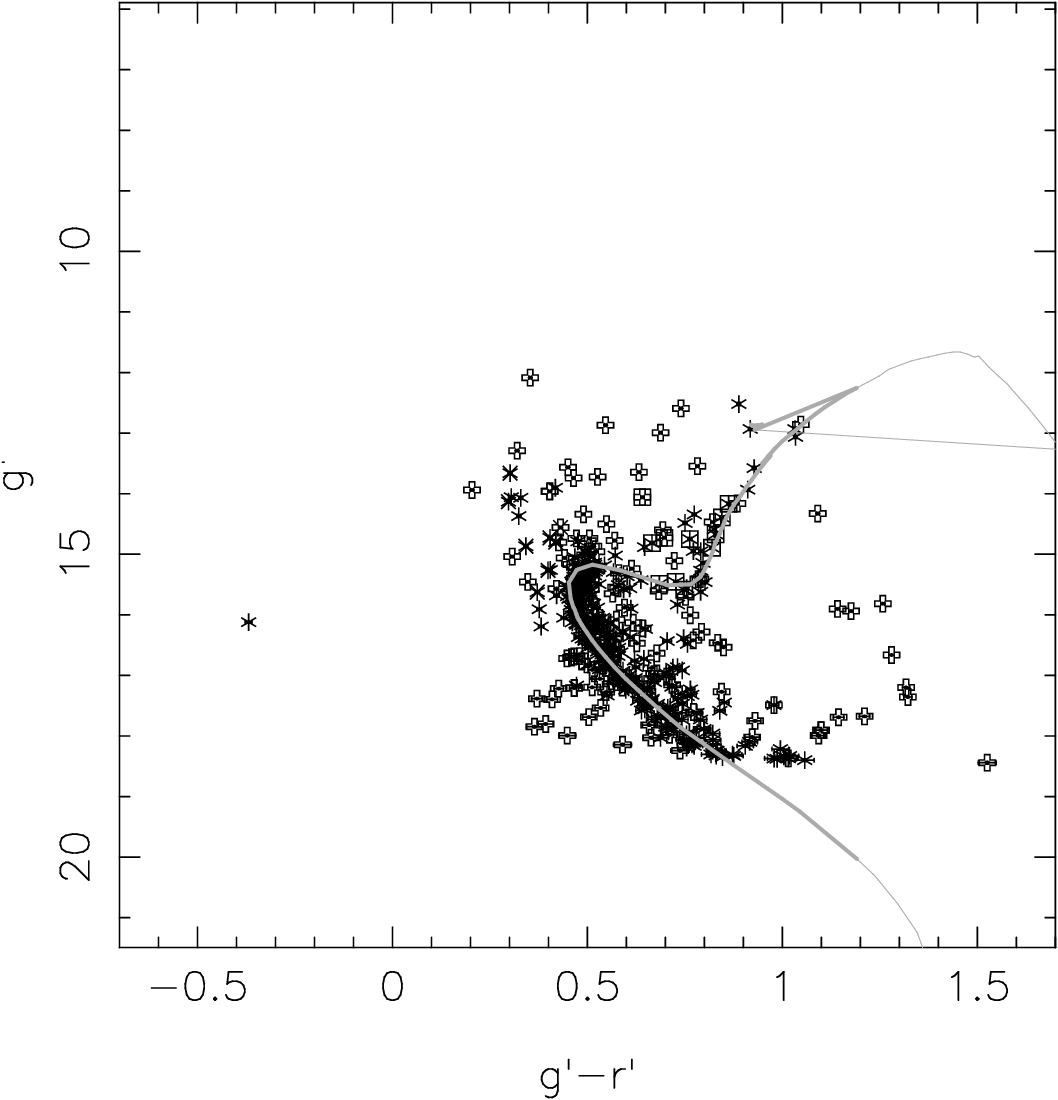}}
\end{picture}
\caption{The observed (not dereddened) color-magnitude diagrams for NGC~188 using 
data from the USNO 1.0 m telescope.  To reduce contamination from
field  stars, only stars with   estimated photon noise  errors in  the
colors less than 0.05~mag  are included (typically stars brighter  than
$g'\approx 15$).  For clarity, we plotted only the 7.1~Gyr isochrone
(distance 1700 pc, $E(B-V)=0.025$, $[Fe/H]=0.0$)
from \citet{Girardi04}. The symbols are the same as in Fig. \ref{webda_fit}
(i.e., asterisks denote proper motion members, thick plus signs indicate 
non-members, squares are red giants, and $\times$ signs are blue stragglers); 
for clarity, the legend key has been omitted from the figures.
}
\label{mosaicCM188}
\end{figure}

We do note that we have
emphasized the $g'$ vs. $(g'-r')$ in our
fits, and that the other diagrams could be better fit by choosing slightly different
parameters.  The slight discrepancies   in  the
best-fit parameters among these diagrams  are within the uncertainties
of the isochrones and  the $AB$ offsets.  Each of    the   color-magnitude diagrams  and    the
color-color diagrams  (Figures \ref{mosaicCM188}  and \ref{mosaicCC188}) has
an  over-plotted  isochrone closest to the one for our best fit
values from above.    The bold
portion of the isochrone represents the region in  which the $ugriz$ to
$u'g'r'i'z'$ transformations are best characterized \citep{Rider04}.

Figures  \ref{mosaicCM188}(a-f)   show  the  full set  of color-magnitude
diagrams available    in the $u'g'r'i'z'$   filter   system for NGC~188.
The $g'$ vs. $(g'-r')$ (Figure \ref{mosaicCM188}b) and the $r'$ vs.  $(g'-r')$ (Figure
\ref{mosaicCM188}c) color-magnitude diagrams are most similar to the more
recognizable $V$ vs.  $(B-V)$ color-magnitude diagrams (see, e.g., \citealt{Karaali05} 
and \citealt{Rodgers06}
for the transformation equations between the $u'g'r'i'z'$ and the Johnson $UBVR_{c}I_{c}$
photometric systems).  
The main sequence and main sequence turn-off are best
defined in the $g'$ vs. $(g'-r')$ diagram.

\placefigure{mosaicCM188}

Figures \ref{mosaicCC188}(a-c) show the three color-color diagrams of the
$u'g'r'i'z'$    filter   system.    The    $(u'-g')$    vs.      $(g'-r')$
(Figure~\ref{mosaicCC188}a)  color-color  diagram is  most similar to the
more  common $(U-B)$  vs.   $(B-V)$ color-color  diagram.

\placefigure{mosaicCC188}

As noted above, the values of distance, age, and reddening $E(B-V)$, which we have determined for NGC~188, are not consistent with the 
{\tt webda} catalog values for this cluster, even within the errors given. The age differs by 75\%, 
the distance by 15\%, and the reddening by 70\% 
($\Delta E(B-V)=0.057$). Such differences in measured values for these cluster parameters are not uncommon (see, Fig. 2 of \citealt{Paunzen06}). 
Furthermore, the values of age and distance agree within the
errors given with the most recent results for NGC~188 (\citealt{Bonatto05}, \citealt{VandenBerg04}, \citealt{Sarajedini99}),
except for the distance presented in \citet{Haroon04}, which is considerably smaller than ours 
and all other modern estimates.
Our value of reddening $E(B-V)$ is close only to the value obtained by \citet{Bonatto05} and \citet{Carraro94}, 
while substantially different from the values presented in \citet{Haroon04}, \citet{VandenBerg04} or \citet{Sarajedini99}.
Since \citet{Bonatto05}, like us, use the Padova isochrones (\citealt{Girardi02}, \citealt{Girardi04}), and since \citet{Haroon04},
\citet{VandenBerg04} and \citet{Sarajedini99} use the VandenBerg \citep{VandenBerg85}, VandenBerg \citep{VandenBerg06} and Yale (\citealt{Chaboyer92}, 
\citealt{Green87}) isochrones, respectively, we suspect the differences are in part due to the isochrones themselves.
In fact, in an important study by \citet{Grocholski03}, they compared several sets of theoretical isochrones and found that 
none of them reproduced the observed multi-band photometry of open clusters in a fully consistent manner, at least over the
magnitude and color range of the non-evolved main sequence. For data in the SDSS filter systems, we are unfortunately 
confined to just the Padova isochrones, as no other sets of theoretical isochrones in the SDSS system have become publicly 
available and current versions of $UBVRI \rightarrow u'g'r'i'z'$ transformation equations are still uncertain at the few percent level.

\section{ Summary and Discussion }

In this third paper of our series, we have analyzed the $u'g'r'i'z'$ photometry
of the open cluster NGC~188. 
Assuming a solar metallicity, we were able to determine a distance of $1700\pm100$ pc, an age of $7.5\pm0.7$ Gyr, 
and a reddening of $E(B-V)=0.025\pm0.005$. 

Our fits turn out to give results which are not
consistent with the values for these parameters found in the {\tt webda} 
on-line open cluster catalog. In particular, our value for $E(B-V)$ is noticeably smaller.
Nevertheless, our results for the age and distance agree well with these parameters'
values from the most recent studies of NGC~188: 7.1~Gyr and 1660 pc (\citealt{Bonatto05}), 
6.8 Gyr and 1685 pc (\citealt{VandenBerg04}), 7 Gyr and 1710 (\citealt{Sarajedini99}).
Although our reddening $E(B-V)$ is neither consistent with 0.087 (\citealt{VandenBerg04}) nor 
0.09 (\citealt{Sarajedini99}), it stays in good agreement with 0.03 (\citealt{Carraro94}) and is
relatively close to 0.00 (\citealt{Bonatto05}). This is very positive since a 
reddening of $E(B-V)=0.087$ stated in \citet{VandenBerg04} was taken from the COBE/DIRBE reddening maps
by \citet{Schlegel98} and thus this would represent an upper limit to the amount of reddening
in the line-of-sight to NGC~188. The \citet{Schlegel98} maps are also very uniform over
the area of NGC~188; therefore, we cannot blame differential reddening over the NGC~188
field as a cause for the different estimates for the reddening.

Summarizing, there is a discrepancy between the {\tt webda} parameters' values for NGC~188 and
our results for this cluster. Nevertheless, the theoretical Padova isochrone on the $g'$ vs. $(g'-r')$
diagram, plotted for our set of values, 
matches the data points very well, precisely restoring the red giant branch. 
Furthermore, our ZAMS theoretical curve fit to the main sequence stars' data points on the 
color-color diagram looks very reliable.

However, there is some probability that the transformation equations 
leading from the $UBV$ set of filters to our $u'g'r'i'z'$ filter system are not precise,
and this could be the reason our value of reddening $E(B-V)$ might not be exact.
In fact, a preliminary version of a new $u'g'r'i'z'$ ZAMS being developed by one of us
(HS; \citealt{Sung06}) tends to indicate a value of $E(B-V)$ more in line with the current {\tt webda}
value of 0.082. This bears further investigation.
In addition, the fact that our results are different than {\tt webda} values might also be caused by
using a different set of theoretical isochrones to fit the data points. The likelihood of 
this scenario becomes even more probable when one considers that the only other recent study 
to give such a low value of $E(B-V)$ for this cluster is that of \citet{Bonatto05}, who also used 
the Padova isochrones (but for $J$ and $H$ 2MASS photometry).
One has to check this issue very carefully.

Since NGC~188 is one of the oldest open clusters known, it is a strategic objective in
understanding the chemical evolution of the Galaxy. It is therefore extremely important
that all the discrepancies between the values of NGC 188's parameters 
be explained.

\acknowledgments

The authors would  like to thank Jeff Pier and Jeff Munn for providing access and support 
for our observations with the US Naval Observatory 1.0 m telescope at Flagstaff Station, Arizona.
We all gratefully acknowledge the support of Leo Girardi for his models. 

We further acknowledge  the {\tt webda} open  cluster database,
operated at the Institute for Astronomy of the University of
Vienna, which made  accessing  available data  for NGC~188
an easy task.

S.S.A. acknowledges support from NSF NVO Grant No. AST-0122449.

B.F. acknowledges  support  from   the   Fermi National   Accelerator
Laboratory Internship  for Undergraduate Physics Majors program during
the summer of 2006.

J.A.S. acknowledges support from NSF Grant No. AST-0098401.

H.S. acknowledges the support of the Korea Science and Engineering Foundation to the 
Astrophysical Research Center for the Structure and Evolution of the Cosmos (ARCSEC) at Sejong University.

Finally, the authors are grateful to the anonymous referee for his/her very useful comments.

This research has made use of the NASA Astrophysics Data System and of
the  Guide Star Catalog  2.2.  The Guide Star Catalogue is a joint project of the Space Telescope Science
Institute and the Osservatorio Astronomico di Torino. Space Telescope Science 
Institute is operated by the Association of Universities for Research in Astronomy,
for the National Aeronautics and Space Administration under contract NAS5-26555.
The participation of the Osservatorio Astronomico di Torino is supported by the
Italian Council for Research in Astronomy. Additional support is provided by
European Southern Observatory, Space Telescope European Coordinating Facility,
the International GEMINI project and the European Space Agency Astrophysics Division.

Funding for the SDSS and SDSS-II has been provided by the Alfred P. Sloan Foundation, 
the Participating Institutions, the National Science Foundation, the U.S. Department of Energy, 
the National Aeronautics and Space Administration, the Japanese Monbukagakusho, 
the Max Planck Society, and the Higher Education Funding Council for England. 
The SDSS Web Site is {\tt http://www.sdss.org/}.

The SDSS is managed by the Astrophysical Research Consortium for the Participating Institutions. 
The Participating Institutions are the American Museum of Natural History, 
Astrophysical Institute Potsdam, University of Basel, Cambridge University, 
Case Western Reserve University, University of Chicago, Drexel University, Fermilab, 
the Institute for Advanced Study, the Japan Participation Group, Johns Hopkins University, 
the Joint Institute for Nuclear Astrophysics, the Kavli Institute for Particle Astrophysics and Cosmology, 
the Korean Scientist Group, the Chinese Academy of Sciences (LAMOST), Los Alamos National Laboratory, 
the Max-Planck-Institute for Astronomy (MPIA), the Max-Planck-Institute for Astrophysics (MPA), 
New Mexico State University, Ohio State University, University of Pittsburgh, University of Portsmouth, 
Princeton University, the United States Naval Observatory, and the University of Washington.


\clearpage

\clearpage

\clearpage

\clearpage

\clearpage

\clearpage

\clearpage

\clearpage

\begin{figure}

\caption{The observed (not dereddened) color-color diagrams for NGC~188 using 
data from the USNO 1.0 m telescope.  To reduce contamination from
field  stars, only stars with   estimated photon noise  errors in  the
colors less than 0.05~mag  are included (typically stars brighter  than
$g'\approx 15$).  For clarity, we plotted only the 7.1~Gyr isochrone 
(distance 1700 pc, $E(B-V)=0.025$, $[Fe/H]=0.0$)
from  \citet{Girardi04}. The symbols are the same as in Fig. \ref{webda_fit}.
}
\label{mosaicCC188}
\end{figure}

\clearpage

\begin{deluxetable}{lccccl}
\rotate
\centering
\tabletypesize{\tiny}
\tablecaption{NGC 188 parameter values\label{table}}
\tablewidth{0pt}
\tablehead{
  \colhead{Reference} &
  \colhead{Distance modulus (distance\tablenotemark{a} \ [pc])} &
  \colhead{Age [Gyr]} &
  \colhead{$E(B-V)$ [mag]} &
  \colhead{$[Fe/H]$} &
  \colhead{Technique} \\
  \colhead{} &
  \colhead{} &
  \colhead{} &
  \colhead{} &
  \colhead{} &
  \colhead{}
}
\startdata

\citealt{Sandage62} &    10.95 (1440)   &   -   &     $0.05\pm0.02$     &  -    &  spectral analysis of classification standards  \\

\citealt{Sandage_2_62}  &     -        &$14-16$ &       -                 & $-0.20$\tablenotemark{b,n}   &  isochrone fits \citep{Hoyle59} \\

\citealt{Demarque64}   &       -      & $9-10$ &           -             & 0.33\tablenotemark{c,n}    &  isochrone fits \citep{Larson64} \\

\citealt{Greenstein64}  &     -        &   -   &    $0.18\pm0.03$       &  -    & spectral analysis of red giants  \\

\citealt{Iben67} &      -       & $11\pm2$ &          -            &   -   &  isochrone fits \\

\citealt{Demarque69}       &	 -     & 7.0  &         -               & 0.70\tablenotemark{d,n}  & isochrone fits \citep{Larson64}\\

\citealt{Eggen69}           & $10.85\pm0.15$ ($1305\pm90$)\tablenotemark{e} & 10 &     0.09              & 0.07  &  $UBV$ photometry \\

\citealt{Sandage69}         &        -     & $8-10$ &           -             &    -   &   isochrone fits    \\
\enddata

\tablenotetext{a}  {calculated from distance modulus after applying a correction of 3.0$\times$$E(B-V)$ 
}
\tablenotetext{b}  {result for $X=0.75$, $Z=0.01$
}
\tablenotetext{c}  {result for $X=0.67$, $Z=0.03$
}
\tablenotetext{d}  {result for $X=0.67$, $Z=0.07$
}
\tablenotetext{e}  {distance modulus based on the Hyades distance modulus of 3.03
}
\tablenotetext{f} {result for $X=0.65$, $Z=0.03$
}
\tablenotetext{g} {result for $X=0.59$, $Z=0.06$
}
\tablenotetext{h} {assumed a solar mixture of heavy elements, i.e. $[M/H]=[Fe/H]$
}
\tablenotetext{i} {result for $X=0.627$, $Z=0.06$
}
\tablenotetext{j} {mean value
}
\tablenotetext{k} {distance molulus based on the Hyades distance modulus of 3.30 
}
\tablenotetext{l} {result for $X=0.70$, $Z=0.03$
}
\tablenotetext{m} {result for $X=0.70$, $Z=0.02$
}
\tablenotetext{n} {transformation equation $X,Z,X_\sun, Z_\sun \rightarrow [Fe/H]$ taken from \citet{Perryman97}, 
value of $(Z/X)_\sun=0.0209$ adopted from \citet{Thoul03}, who state it after \citet{Grevesse02}; also assumed a solar mixture of heavy elements
}
\tablenotetext{p} {adopted $E(B-V)$ for the giants
}
\tablenotetext{q} {the same result previously in \citet{Worthey02}
}
\tablenotetext{r} {$E(B-V)$ taken from \citet{Schlegel98} dust maps
}
\tablenotetext{s} {explored range of metallicity: -0.12 $\leq$ $[Fe/H]$ $\leq$ 0.01, corresponding age and distance modulus 
varied from 5.9 to 8.1 Gyr and from 11.22 to 11.54 mag, respectively, which results in distance from 1555 to 1800 pc
}

\end{deluxetable}

\clearpage

\begin{deluxetable}{lcccccccl}
\centering
\tablecaption{Observing Circumstances\label{obs}}
\tablehead{
  \colhead{Date} &
  \colhead{MJD} &
  \colhead{Airmass} &
  \multicolumn{5}{c}{Exposure (sec)} &
  \colhead{Comments} \\
  \colhead{} &
  \colhead{} &
  \colhead{} &
  \colhead{ $r'$ } &
  \colhead{ $g'$ } &
  \colhead{ $u'$ } &
  \colhead{ $i'$ } &
  \colhead{ $z'$ } &
  \colhead{}
}
\startdata
2002 Nov 05 & 52583 &  1.558  & 10  & 10  & 70  & 10  & 15  & Non-Photometric \\
            &  &  1.559   & 60  & 60 & 420 &  60 & 90 & Photometric \\
            &  &  1.563   & 240  & 240  & 1680  & 240  & 360  & Photometric \\
            &  &  1.578  & 10  & 10  & 70  & 10  & 15  & Non-Photometric \\  
\enddata
\end{deluxetable}

\clearpage

\begin{deluxetable}{lccccccc}
\tablecaption{Night Characterization Coefficients\tablenotemark{a} \ for MJD 52583\label{nightChar}}
\tablewidth{0pt}
\tablehead{
 
  \colhead{Filter} &
  \colhead{Zeropoint (a)} &
  \colhead{1st-Order Ext. (k)} &
  \colhead{Std. rms} &
  \colhead{\# Std.} \\
  \colhead{}  &
  \colhead{}  &
  \colhead{}  &
  \colhead{(mag)}  &
  \colhead{stars}  \\
  \colhead{(1)}  &
  \colhead{(2)}  &
  \colhead{(3)}  &
  \colhead{(4)}  &
  \colhead{(5)}  
}
\startdata                                                                        

 $u'$ & -20.088$\pm$0.032  & 0.470$\pm$0.022 & 0.027 & 22 \\
 $g'$ & -21.658$\pm$0.011  & 0.152$\pm$0.008 & 0.012 & 27 \\
 $r'$ & -21.502$\pm$0.011  & 0.081$\pm$0.007 & 0.012 & 27 \\
 $i'$ & -21.083$\pm$0.016  & 0.036$\pm$0.011 & 0.016 & 26 \\
 $z'$ & -19.928$\pm$0.026  & 0.005$\pm$0.017 & 0.026 & 27 \\
\enddata

\tablenotetext{a} {The values for the second order extinction term ($c$) 
coefficients were set  to $-2.1 \times 10^{-2}$, $-1.6\times 10^{-2}$,
$-4.0 \times 10^{-3}$, $6.0 \times 10^{-3}$  and $3.0 \times 10^{-3}$,
for  the $u'$,$g'$,$r'$,$i'$ and  $z'$  respectively. 
The values for the Instr. Color ($b$) were all set to zero.
These values are
those determined in \citet{Smith02}.}

\end{deluxetable}

\clearpage

\begin{deluxetable}{l@{ }c@{ }c@{ }cc@{ }c@{ }c@{ }c@{ }cc@{ }c@{ }c@{ }c@{ }cc@{ }c@{ }c@{ }c@{ }c@{ }c@{ }cl}
\rotate
\tabletypesize{\tiny}
\tablecaption{Available Data from the USNO 1.0~m Telescope for NGC~188\label{dataUSNO}}
\tablewidth{0pt}
\tablehead{
  \colhead{Star ID} &
  \colhead{WEBDA ID} &
  \colhead{RA} &
  \colhead{DEC} &
  \multicolumn{5}{c}{Magnitude\tablenotemark{*}} &	
  \multicolumn{5}{c}{Magnitude Error} &
  \multicolumn{5}{c}{Flag} &
  \colhead{Membership} &
  \colhead{Comments} \\	
  \colhead{} &
 \colhead{} &	 
  \colhead{} &
  \colhead{} &
  \colhead{ $u'$ } &
  \colhead{ $g'$ } &
  \colhead{ $r'$ } &
  \colhead{ $i'$ } &
  \colhead{ $z'$ } &
 \colhead{ $u'$ } &
  \colhead{ $g'$ } &
  \colhead{ $r'$ } &
  \colhead{ $i'$ } &
  \colhead{ $z'$ } &
 \colhead{ $u'$ } &
  \colhead{ $g'$ } &
  \colhead{ $r'$ } &
  \colhead{ $i'$ } &
  \colhead{ $z'$ } &
  \colhead{Probability} & 
  \colhead{} \\
 \colhead{1} &
 \colhead{2} &
 \colhead{3} &
 \colhead{4} &
 \colhead{5} &
 \colhead{6} &
 \colhead{7} &
 \colhead{8} & 
 \colhead{9} &
 \colhead{10} &
 \colhead{11} &
 \colhead{12} &
 \colhead{13} &
 \colhead{14} &
 \colhead{15} &
 \colhead{16} &
 \colhead{17} &
 \colhead{18} &
 \colhead{19} &
 \colhead{20} & 
 \colhead{21} 
}
\startdata
1 & -1 & 00:47:57.12 & 85:14:56. & 14.858 & 12.520 & 11.632 & \nodata & 11.087 & 0.002 & 0.003 & 0.002 & \nodata & 0.015 & 0 & 0 & 0 & 1 & 0 & 96 & d \\
2 & -1 & 00:46:19.43 & 85:13:09. & 13.244 & 12.084 & 11.731 & 11.623 & 11.587 & 0.002 & 0.003 & 0.002 & 0.003 & 0.015 & 0 & 0 & 0 & 0 & 0 & 0 \\
3 & -1 & 00:47:56.30 & 85:11:34. & 15.583 & 12.860 & 11.812 & 11.433 & 11.158 & 0.003 & 0.003 & 0.002 & 0.003 & 0.015 & 0 & 0 & 0 & 0 & 0 & 0 \\
4 & -1 & 00:50:44.06 & 85:17:42. & 14.559 & 12.592 & 11.852 & 11.541 & 11.358 & 0.005 & 0.003 & 0.002 & 0.003 & 0.015 & 0 & 0 & 0 & 0 & 0 & 0 \\
5 & -1 & 00:44:52.56 & 85:14:05. & 15.868 & 12.934 & 11.901 & 11.508 & 11.263 & 0.003 & 0.003 & 0.002 & 0.003 & 0.015 & 0 & 0 & 0 & 0 & 0 & 98 \\
6 & -1 & 00:46:59.77 & 85:13:16. & 15.575 & 12.931 & 12.014 & 11.683 & 11.469 & 0.003 & 0.003 & 0.002 & 0.003 & 0.015 & 0 & 0 & 0 & 0 & 0 & 98 \\
7 & -1 & 00:45:51.07 & 85:18:08. & 15.986 & 13.062 & 12.028 & 11.632 & 11.384 & 0.003 & 0.003 & 0.002 & 0.003 & 0.015 & 0 & 0 & 0 & 0 & 0 & 96 \\
8 & -1 & 00:51:55.35 & 85:16:19. & 14.839 & 12.994 & 12.306 & 12.055 & 11.938 & 0.002 & 0.003 & 0.002 & 0.003 & 0.014 & 0 & 0 & 0 & 0 & 0 & 0 \\
\enddata

\end{deluxetable}

\end{document}